\documentclass[twocolumn,showpacs,preprintnumbers,amsmath,amssymb]{revtex4}

\usepackage{graphicx}% Include figure files
\usepackage{dcolumn}% Align table columns on decimal point
\usepackage{bm}% bold math
\usepackage{epsfig}% old epsfig
\usepackage{psfrag}%psfrag labels

\usepackage{amssymb}
\usepackage{epstopdf}

\def\eq#1{Eq.~(\ref{#1})}

\newcommand{\half}{{ \frac{\mbox{\scriptsize 1}}{\mbox{\scriptsize 2}}}}
\newcommand{\quarter}{{ \frac{\mbox{\scriptsize 1}}{\mbox{\scriptsize 4}}}}
\newcommand{\eighth}{{ \frac{\mbox{\scriptsize 1}}{\mbox{\scriptsize 8}}}}
\newcommand{\Tr}{\mbox{Tr}}

\begin{document}

%\preprint{}

\title{Non-Boltzmann behaviour in models of interacting neutrinos}

\author{Bruce H J McKellar}%
 \email{bhjm@unimelb.edu.au}
\author{Ivona Okuniewicz} \email{ivona.okuniewicz@gmail.com}
\author{James Quach} \email{jamesq@unimelb.edu.au}
\affiliation{%
School of Physics, University of Melbourne, Victoria Australia 3010
}%

\date{18 March 2009}

\begin{abstract}
We reconsider the question of the relative importance of single particle effects and correlations in the solvable interacting neutrino models introduced  by  Friedland and Lunardini and by Bell, Rawlinson and Sawyer.  We show, by an exact calculation, that the two particle correlations are not ``small'', and that they dominate the time evolution in these models,  in spite of indications to the contrary from the rate of equilibration.  This result holds even after the  model in generalized from the original $2$ flavor case to $N$ flavors. The failure of the Boltzmann single particle approximation in this model is tentatively attributed to the simplicity of the model,  in particular to the assumption that all neutrinos in the initial state are in flavor eigenstates.\end{abstract}

\pacs{14.60Pq, 05.30.-d}

\maketitle

\section{\label{sec:intro}Introduction}

This paper is an extension of the work of BMcK and IO in collaboration with Alexander Friedland\cite{Friedland:2006ke}.  Preliminary accounts of some of the results have appeared in the PhD thesis of IO\cite{Okuniewicz:2006th} and the BSc honours report of JQ\cite{Quach:2009vh}.  Details of the  calculations omitted here, can be found in the thesis and the report.  An analysis of the two flavor case was briefly reported earlier\cite{McKellar:2008yc, McKellar:2009ij}.

In the early 1990s a number of authors, including Thomson and McKellar\cite{Thomson:1990th,Thomson:1990jj,Thomson:1990hj,Thomson:1990hk, McKellar:1992ce, McKellar:1992ja}, Pantelone\cite{Pantaleone:1992eq, Pantaleone:1992xh}, and Sigl and Rafelt\cite{Sigl:1992fn},  derived neutrino kinetic equations which took into account neutrino mixing and interactions.  They were a set of coupled equations involving the number densities (or the single particle density matrices) for the different species of neutrinos and anti-neutrinos.  These kinetic equations were derived from the quantum statistical mechanical equations for the many particle density matrix by a process similar to the derivation of the quantum Boltzmannn equation.  In particluar it was assumed that many body correlations do not play a significant role in the time evolution of the single particle density matrix, although, as we discuss below,  this is not the only assumption in the derivation of the neutrino kinetic equations (or the quantum Boltzmannn equation).  Neutrino kinetic equations have proved to be very useful in studies of the influence of neutrinos on the evolution of the early universe\cite{Foot:1995qk, Foot:1996qc, Bell:1998ds, Dolgov:2002wy}, and are now in use in the study of supernovae\cite{Dasgupta:2007ws}.

In 2003, Friedland and Lunardini\cite{Friedland:2003dv},  and Bell, Rawlinson and Sawyer\cite{Bell:2003mg} re-examined the validity of the single-particle approximation inherent in these kinetic equations.  Part of the motivation was a concern that the entanglement induced by the interactions could make correlations so important that the single particle approximation could be invalidated.  They studied a simplified model which allowed exact solutions of the many body problem, and examined the rate of equilibration of the system.   This discussion was extended by Friedland and Lunardini\cite{Friedland:2003eh} and by Friedland, McKellar and Okuniewicz\cite{Friedland:2006ke}.  The crucial point in the discussion is the observation that the incoherent equilibration time and the coherent equilibration times scale with the number of particles in very different ways:
\begin{eqnarray}
t ^{\mbox{{\scriptsize inc}}}_{\mbox{\scriptsize eq}} & \propto &  1/(g\sqrt{N}), \\
t ^{\mbox{\scriptsize coh}}_{\mbox{\scriptsize eq}} & \propto &  1/(gN).
\end{eqnarray}
Here $g$ is a suitably normalized interaction strength.  In a large system coherent equilibration is much faster than incoherent equilibration, and thus the equilibration timescale of the many body system can be used as a proxy to determine the validity or otherwise of the single particle approximation, in as much as one does not have coherent equilibration in such an approximation.  Bell, Rawlinson and Sawyer found a coherent timescale in their numerical modeling, but Friedland and Lunardini, and  Friedland, McKellar and Okuniewicz found incoherent timescales.  While these three papers studied systems that differed in detail, they   lead to the conclusion that, at least in many circumstances, the single particle approximation, and thus the neutrino kinetic equations or the quantum Boltzmann equation, are valid.

In this paper we demonstrate that this conclusion  is  false.  The flaw in the logic is that while coherent equilibration timescales are an indication that correlations are important, incoherent equilbration timescales do not necessarily imply that correlations are insignificant. 

Because  the model of Friedland, McKellar and Okuniewicz is exactly solvable, we can explicitly calculate the two body correlations in this model, and we find that, in the case when the number of flavors,  $\mathcal{N}$, is two, they are significant.  Moreover we verify that the time evolution of the system is determined completely and correctly by the two body correlations through the BBGKY equations.  We go on to demonstrate that the one particle approximation, \emph{i.e.} the quantum Boltzmannn equation, with the initial condition that all particles are in flavor eigenstates, leads to the erroneous conclusion that the single particle density matrix is constant in time. We show that this final conclusion remains valid in the $\mathcal{N}$ flavor generalization of the model.

In section II we review the model of Friedland, McKellar and Okuniewicz, including the $\mathcal{N}$ flavor generalization to $SU(\mathcal{N})$ and we then organize the paper along the above lines --- section III shows that the two body correlations are significant, section IV demonstrates that the two body correlations and the BBGKY equations correctly determine the time evolution of the one particle density matrix, and section V shows how the Boltzmannn equation fails to determine the time evolution of the single particle density matrix.  Finally in section VI we review our results and discuss their implications.

\section{\label{sec:model}The model of Friedland, McKellar and Okuniewicz}
 Friedland, McKellar and Okuniewicz used the model of Friedland and Lunardini, and generalized it to obtain the solution to a system of neutrinos with unequal numbers of flavors.  So that the system could be easily solved explicitly and exactly it was restricted to two neutrino types and the spatial or momentum degrees of freedom were ignored.  We extend the model to more than two neutrino flavors, but the $SU(\mathcal{N})$ analogue of the 6-j coefficients of SU(2) are not so well studied, making the explicit analysis more difficult to complete. Nevertheless, some general results can be obtained for the $\mathcal{N}$ flavor generalization.  The advantage of the model is that is allows an exact investigation of the dynamics of flavor conversion. 

With two neutrino flavors we could regard the neutrino system as equivalent to the up-down quark system (or the proton neutron system) and use the language of isospin in talking about the system.  While the the earlier papers used the language of spin, because we are considering the many flavor generalisation from time to time in this paper, we will use the flavor symmetry language, and for $\mathcal{N}=2$, the language of isospin.

Neglecting the spatial degrees of freedom  (and thus the Dirac matrices), the weak interaction of a pair of neutrinos inducing flavor mixing in the $\mathcal{N}$ flavor case is 
\begin{equation}
H_{\mbox{\scriptsize W}} = g  \left(
\delta_{\iota\kappa}\delta_{\lambda\mu} + \delta_{\iota\mu}\delta_{\lambda\kappa} \right),
\label{eqn:hweak}
\end{equation}
where $g$ is a coupling constant.  This Hamiltonian acts on a product of two $\mathcal{N}$-dimensional vectors, the neutrino wavefunctions in a flavor-space representation.
We introduce the $\mathcal{N}\times \mathcal{N}$ Gell-Mann matrices\cite{gm89},\footnote{The $\mathcal{N} = 3$ Gell-Mann matrices were introduced into the description of the three flavor neutrino problem by Thomson and McKellar\cite{Thomson:1990hk}, and are now widely used for the purpose, see eg ref.\cite{Dasgupta:2007ws}}  $\lambda_j, j \in \{0, 1, \ldots ,\mathcal{N}^2-1\}$, with the normalization
\begin{equation}
\mbox{Tr}(\lambda_j \lambda_k) = 2 \delta_{jk}
\end{equation}
and the conventions that 
\begin{equation}
\lambda_0 = \sqrt{2/\mathcal{N}} \mathbf1_{\mathcal{N}}, \quad \Tr{\lambda_j} = 0\;\; \mbox{for}\;\; j \ne 0,
\end{equation}
where $ \mathbf1_{\mathcal{N}}$ is the $\mathcal{N} \times \mathcal{N}$ unit matrix. The diagonal matrices are, for $n \in \{2,\ldots, \mathcal{N}\}$ 
\begin{equation}
\lambda_{n^2 - 1} = \sqrt{\frac{2}{n(n-1)}}\mbox{diag}\{1,1,\ldots,1,-(n-1),0,\ldots,0\}
\end{equation} 
with $n-1$ ones and $\mathcal{N} - n$ zeros along the diagonal.  For the non-diagonal matrices we choose a notation which simply extends the Gell-Mann notation for $\mathcal{N} = 3$.  For $n \in \{1, \ldots, \mathcal{N}\}$, and $r \in \{1, \ldots, n\}$,
\begin{eqnarray}
\{\lambda_{n^2-1 + 2r -1}\}_{i,j}  = \delta_{i,r}\delta_{j,n+1} + \delta_{i,n+1}\delta_{j,r}, \\
\{\lambda_{n^2-1 + 2r}\}_{i,j}  = -i\left(\delta_{i,r}\delta_{j,n+1} - \delta_{i,n+1}\delta_{j,r} \right).
\end{eqnarray}

Using the identity
\begin{equation}
\delta_{\iota\mu}\delta_{\lambda\kappa} =\left(\frac{1}{\mathcal{N}}\delta_{\iota\kappa}\delta_{\lambda\mu} +\frac{1}{2} \sum_{j = 1}^{\mathcal{N}^2-1}(\lambda_j)_{\iota\kappa}(\lambda_j)_{\lambda\mu}\right)
\end{equation}
 the pair Hamiltonian can be rewritten as
\begin{eqnarray}
%\begin{split}
H_{\mbox{\scriptsize W}}(ij)  & =  & g \left(\frac{\mathcal{N}+1}{\mathcal{N}}+\frac{\boldsymbol{\lambda_{i}}\cdot\boldsymbol{\lambda_{j}}}{2} \right) \\
&= & g\left[\left(\mathbf{F}_i+\mathbf{F}_j\right)^2 -\frac{(\mathcal{N}-2)(\mathcal{N} + 1)}{\mathcal{N}}\right]. 
\label{eqn:spinInteraction}
%\end{split}
\end{eqnarray}
The ``F-spin'', $\mathbf{F} = \boldsymbol{\lambda}/2$,   is the $SU(\mathcal{N})$ equivalent of isospin.  The interaction Hamiltonian for N neutrinos is then
\begin{equation}
\label{eqn:interactionHamiltonian}
H =g\sum_{i=1}^{N-1}\sum_{j=i+1}^{N} \left(\frac{\mathcal{N}+1}{\mathcal{N}}+\mathbf{F}_{i}\cdot\mathbf{F}_{j} \right)
\end{equation}
which is in turn a linear function of the square of the total F-spin,
\begin{equation}
\mathbf{F} = \sum_{j=1}^{N} \mathbf{F}_j,
\end{equation}
the possible values of which are completely determined by the initial configuration of the neutrino flavors.  As $\mathbf{F}^2$ is a constant of the motion, the interaction Hamiltonian is completely determined by the eigenstates and eigenvalues of the system, which are well defined.

In the particular case $\mathcal{N} = 2$, the eigenvalues are completely determined by the the total isospin $T$ and the number of particles $N$,
\begin{equation}
E_{T}=[g T(T+1)+g\frac{3}{4}N(N-2)].
\end{equation}
The eigenstates $\Psi_{T, M, \alpha}$ are not uniquely determined by $T,M$.  One also needs to specify $M$, the projection of the total isospin on the quantization axis, and additional quantum numbers $\alpha$, which specify the different ways that states of total isospin $T$ can be constructed.  Note that, because we are ignoring the position (or momentum) variables, we are unable to generate states which obey Fermi statistics --- any two isospin up particles are necessarily identical and thus antisymmetrizing them produces a null state, as required by the Pauli principle.
Take as an initial state a state with  definite numbers of isospin up and isospin down particles, labelled by $U$ and $D$ respectively,
\begin{equation}
\label{eqn:intialState}
\left|\Psi(t=0)\right\rangle = \left|T_{U} M_{U}\right\rangle\otimes\left|T_{D} M_{D}\right\rangle=\left|T_{U} T_{D};M_{U} M_{D}\right\rangle.
\end{equation}

Because we know the eigenstates and the eigenvalues of the Hamiltonian of \eq{eqn:interactionHamiltonian}, we can easily write down how this state evolves with time by expanding the initial state in terms of basis states with definite  total isospin and its projection, $(T,M)$, which thus have a well defined time dependence:
\begin{equation}
\label{eqn:totalBasisState}
\begin{split}
\left|\Psi(t)\right\rangle & =\sum_{T}e^{-itE(T)}\left\langle T,M\right|\left.M_{U} M_{D};T_{U} T_{D}\right\rangle \\
& \quad \left |T_{U} T_{D};T M\right\rangle.
\end{split}
\end{equation}
To get the one body density matrix for particle $1$, which we denote as $\rho^{(1)}_{1}$ \footnote{the superscript $^{(1)}$ indicates that the density matrix refers to just one particle, and the subscript $_1$ indicates that the one particle is particle number $1$}, we need to single out that particular particle. To do this we decouple it from the rest, writing for the initial
\begin{equation}
\left|T_{U} T_{D}\right\rangle=\left|(I_{1} k_{U})T_{U},T_{D}\right\rangle
\end{equation}
where $I_{1}=1/2$ is the isospin of the one particle,
$k_{U} = T_U - 1/2$ is the total angular momentum of the rest of the spin-up particles.

Note that we have specified that the particle we are singling out is initially spin-up.  Of course, an analogous derivation can be done for a particle that is initially spin-down.

To construct the  density matrix of the total system, we use the basis $ \left|I_1I;m_1m \right\rangle$ in which $k_U$ and $T_D$ are coupled to isospin $I$, with projection $m$, but to get the time dependence we must relate this to states with a fixed total isospin $T$.  The result is
\begin{equation}
\begin{split}
		\rho(t)&=\left|\Psi(t)\right\rangle\left\langle\Psi(t)\right|\\
		&= \sum_{\substack{T,T',I,I',I_{1},I'_{1}\\m,m',m_{1},m'_{1}}}\exp\{-itg[T(T+1)-T'(T'+1)]\}\\
		& \quad (2I_n+1) \sqrt{(2I+1)(2I'+1)}\\
		&\quad\left\langle T_{U} T_{D}; T M \right|\left. M_{U} M_{D}\right\rangle \left\langle T_{U} T_{D}; T' M \right|\left. M_{U} M_{D} \right\rangle\\
		&\quad\left\{
			\begin{array}{ccc}
				1/2 & T_{U}-1/2 & T_{U}\\
				T_{D} & T & I
			\end{array}
		\right\}
		\left\{
			\begin{array}{ccc}
				1/2 & T_{U}-1/2 & T_{U} \\
				T_{D} & T' & I'
			\end{array}
		\right\}\\
		&\quad \left|I_1I;m_1m \right\rangle \left\langle I_1'I';m_1'm'\right|.
\end{split}
\end{equation}
To obtain the one-body density matrix, we take the trace over the other particles to get
\begin{equation}
\label{oneBodyDensityMatrixUp}
\begin{split}
		\rho^{(1)}_1(t)&=\sum_{\substack{T,T',I,I_{1},I'_{1}\\m,m_{1},m'_{1}}}\exp\{-itg[T(T+1)-T'(T'+1)]\}\\
		& \quad (2I_1+1)(2I+1) \\
		&\quad \left\langle T_{U} T_{D}; T M \right|\left.  M_{U} M_{D}\right\rangle \left\langle T_{U} T_{D}; T' M \right|\left. M_{U} M_{D} \right\rangle \\
		&\quad \left\langle I_1 I; T M | m_1 m\right\rangle \left\langle I_1 I; T' M \right|\left. m'_1 m\right\rangle\\
		&\quad\left\{
			\begin{array}{ccc}
				1/2 & T_{U}-1/2 & T_{U} \\
				T_{D} & T & I
			\end{array}
		\right\}
		\left\{
			\begin{array}{ccc}
				1/2 & T_{U}-1/2 & T_{U} \\
				T_{D} & T' & I
			\end{array}
		\right\}\\
		&\quad \left| I_1 m_1\right\rangle \left\langle I_1 m'_1\right|.
\end{split}
\end{equation}
\eq{oneBodyDensityMatrixUp} is for a particle that initially has spin up. For a particle that initially has spin down, one need only interchange the subscripts $U$ with $D$.  This one body density matrix can now be used to calculate the probability that the particle has spin up at time $t$, or the expectation value of the spin at time $t$.  It was the basic tool used by Friedland, McKellar and Okuniewicz to calculate the time evolution of single particle probabilities.

We wish to calculate two body correlation functions, and also test the quantum BBGKY and Boltzmannn equations.  For these purposes we also require the two body density matrix, which is just as straightforward to calculate.

 If the two particles initially have the same isospin projection, the derivation follows the above, with the difference being that one must decouple two spin-up particles instead of one. The final result is that one need only make the following changes to \eq{oneBodyDensityMatrixUp},
\begin{itemize}
\item replace $I_1= 1/2$ with $I_{12} = 1$ where $I_{12}$ is the summation of the angular momentum of the two particles \emph{i.e.}  $\vec{I}_{12}=\vec{I}_1+\vec{I}_2$  (because both particles have isospin up, the pair must have isospin 1),
\item in the $6j$ symbols replace $T_{U} - 1/2$ with $T_{U} - 1$,
\item replace $m_1$ with $m_{12}$ where $m_{12}=-I_{12},-I_{12}+1,...,I_{12}-1,I_{12}$ is the isospin projection of the pair of particles,
\end{itemize}
in order to obtain the two body density matrix.  When both isospins are initially up the result is
\begin{equation}
\label{twoBodyDensityMatrixUpUp}
\begin{split}
		\rho^{(2)}_{12}(t)&=\sum_{\substack{T,T',\\m,m_{12},m'_{12}}}\exp\{-itg[T(T+1)-T'(T'+1)]\}\\
		&\quad (2I_{12}+1)(2I+1) \\
		&\quad\left\langle T_{U} T_{D}; T M | M_{U} M_{D}\right\rangle  \left\langle T_{U} T_{D}; T' M | M_{U} M_{D}\right\rangle \\
		&\quad\left\langle I_{12} I; T M | m_{12} m\right\rangle  \left\langle I_{12} I; T' M | m'_{12} m\right\rangle \\
		&\quad\left\{
			\begin{array}{ccc}
				1 & T_{U}-1 & T_{U} \\
				T_{D} & T & I
			\end{array}
		\right\}
		\left\{
			\begin{array}{ccc}
				1 & T_{U}-1 & T_{U} \\
				T_{D} & T' & I
			\end{array}
		\right\}\\
		&\quad \left| I_{12} m_{12}\right\rangle \left\langle  I_{12} m'_{12} \right|.
\end{split}
\end{equation}

For particles that initially have opposite spins the analysis is a little different.  In this case, decouple one particle from the isospin up system, and one from the isospin-down system.   The $N_U - 1$ remaining isospin up particles have a total isospin $k_U = T_U - 1/2$, which couples with $j_1$ to give the isospin $T_U$.  The isospin down particles are similarly recoupled.   Then the single particle isospins $I_1$ and $I_2$ are coupled to an isospin $I_{12}$, $k_U$ and $k_D$ are coupled to the isospin $I$, and $I_{12}$ and $I$ are coupled to $T$.
Standard angular momentum manipulations then allow us to construct the two body density matrix for this case, which now involves 9-j symbols:
\begin{equation}
\label{twoBodyDensityMatrix}
	\begin{split}
		\rho_{12}(t)&=\sum_{\substack{T,T',I,I_{12},I'_{12}\\m,m_{12},m'_{12}}}\exp\{-itg[T(T+1)-T'(T'+1)]\}\\
		&\quad(2T_{U}+1)(2T_{D}+1)(2I+1)\\
		& \quad \times [(2I_{12}+1)(2I'_{12}+1)]^{\frac{1}{2}}\\
		&\quad\left\langle T_{U} T_{D}; T M | M_{U} M_{D}\right\rangle  \left\langle T_{U} T_{D}; T' M | M_{U} M_{D}\right\rangle \\
		&\quad\left\langle I_{12} I; m_{12} m |T M\right\rangle  \left\langle I'_{12} I; m'_{12} m |T' M\right\rangle \\
		&\quad\left\{
			\begin{array}{ccc}
				I_1 & k_U & T_{U} \\
				I_2 & k_D & T_{D}\\
				I_{12} & I & T\\
			\end{array}
		\right\}
		\left\{
			\begin{array}{ccc}
				I_1 & k_U & T_{U} \\
				I_2 & k_D & T_{D}\\
				I_{12} & I & T'\\
			\end{array}
		\right\}\\
	&\quad \left|I_{12} m_{12}\right\rangle \left\langle I'_{12} m'_{12}\right|.\\
	\end{split}
\end{equation} 

With these basic ingredients we can proceed to obtain the results outlined in the introduction.  First we calculate the two body correlations.

\section{\label{sec:corr}Two particle correlations}
With expressions for the one and two particle density matrix we can easily calculate the two body correlation function, 
\begin{equation}
\Gamma = \langle t_1 t_2 \rangle - \langle t_1 \rangle\langle t_2\rangle \label{corrfn}
\end{equation}
where $\langle t_i \rangle$ is the expectation value of the isospin $t_i$, of particle $i$, and so on.

It may be helpful to emphasize that, since there is no momentum or spatial dependence in this model, the usual use of the correlation function to determine correlations between particles at different locations does not apply.  $\Gamma$ simply measures the correlation between isospins, but its time dependence is of interest.

Expectation values may be obtained from density matrices in the usual way, \emph{e.g.}
\begin{equation}
\langle t_1 t_2 \rangle = \mbox{Tr}\left\{ \rho_{12}\left|m_1m_2\rangle \langle m_1m_2 \right| \right\},
\end{equation}
or equivalently by first using the density matrix to obtain the probability $P(m_1,m_2)$ that the isospin projections of the two particles are $m_1$ and $m_2$, and then calculating
\begin{equation}
\langle t_1 t_2 \rangle = \sum_{m_1,m_2} m_1m_2 P(m_1,m_2).
\end{equation}
Since the details of these calculations are given in the PhD thesis of Okuniewicz, we simply give the results here.

\subsection{The correlation function when both particles are initially in isospin-up states}
\label{sec:Gammaupup}

In the case where both particles are initially in the isospin-up state,  \eq{twoBodyDensityMatrixUpUp} applies.  The final expression for the correlation function is 
rather long and not particularly instructive.  We give some results using it, allowing the interested reader to  obtain the details  from reference \cite{Okuniewicz:2006th}.  

\begin{figure}[h]
\psfrag{G}{$\Gamma$}
\psfrag{t}{$\tau$}
\begin{center}
\fbox{\epsfig{file=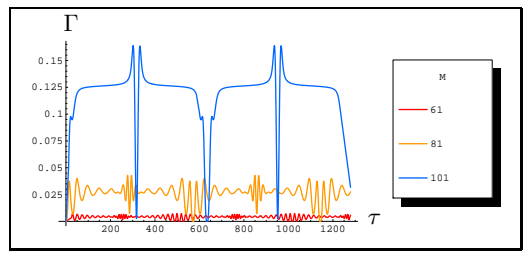,scale=0.4}}
\caption{A plot of the correlation function, $\Gamma$, using  the initial conditon  in which both particle 1 and particle 2 have isospin-up.  In these plots $N=101$, M is varied such that $N\geq M$ .  The time is scaled as usual, $\tau =gt(N+M)$.  Note that the correlation function is positive for all cases.  For the case $N=M$ the glitches in the plot are not real, rather they are an artifact of mathematica graphics}
\label{fig:up a}
\end{center}
\end{figure}

\begin{figure}[h]
\psfrag{G}{$\Gamma$}
\psfrag{t}{$\tau$}
\begin{center}
\fbox{\epsfig{file=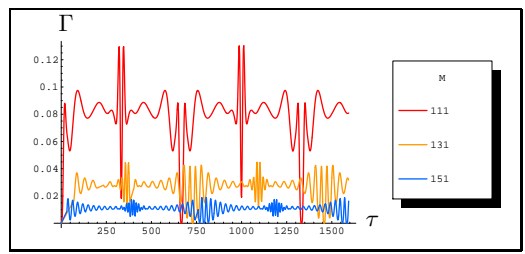,scale=0.4}}
\caption{A plot of the correlation function, $\Gamma$, using  the initial condition in which both particle 1 and particle 2 have isospin-up.  In these plots $N=101$ and M is varied such that $N < M$ .  The time is scaled as usual, $\tau =gt(N+M)$.  Note that the correlation function is positive for all cases.}
\label{fig:up b}
\end{center}
\end{figure}

 To analyse the correlation function  we have plotted it as a function of the time for various numbers of isospin up particles, N, and isospin down particles, M.  A subset of these simulations is shown in figure \ref{fig:up a} and figure \ref{fig:up b} for which $N=101$ and M is varied. 

The correlation function shows the same features as those found by Friedland, McKellar and Okuniewicz for the probability of one of the initial isospin up particles remaining in the isospin up state.  Namely, 
\begin{itemize}
\item When $N\sim M$, the system comes to equilibrium after some time. This features can be seen for the case N=101, M=101.  Just as in the study of the equilibration of the single isospin by Friedland, McKellar and Okuniewicz, the equilibration time is incoherent.
\item When $N-M$ is large the correlation function exhibits oscillations but the amplitude is small.
\item The correlation function is periodic on large time scales, with the same period as before.
\end{itemize}

There are two new features to note:
\begin{itemize}
\item the correlation is always positive,
\item when the equilibration time is clearly incoherent ($N \sim M$), the correlation is of order 10\%, which is not particularly small.
\end{itemize}
We defer discussion of these features until we have presented the results for the case where the isospins are initially anti-parallel.

\subsection{The correlation function when both particles are initially in anti-parallel isospin states}
\label{sec:Gammaupdown}

\begin{figure}[h]
\psfrag{G}{$\Gamma$}
\psfrag{t}{$\tau$}
\begin{center}
\fbox{\epsfig{file=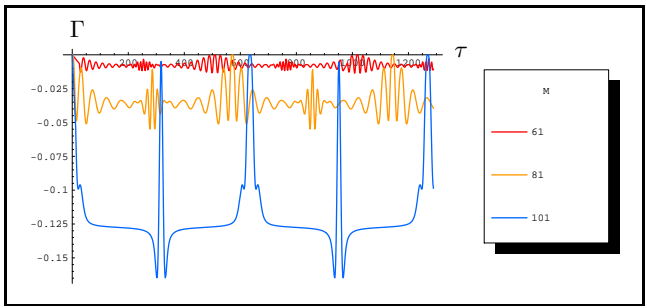,scale=0.4}}
\caption{A plot of the correlation function, $\Gamma$, using the initial conditon  in which particle 1 has isospin-up and particle 2 has isospin-down.  In these plots $N=101$, M is varied such that $N\geq M$ .  The time is scaled as usual, $\tau =gt(N+M)$.  Note that the correlation function is negative for all cases.  For the case $N=M$ the glitches in the plot are not real, rather they are an artifact of mathematica graphics}
\label{fig:down a}
\end{center}
\end{figure}

\begin{figure}[h]
\psfrag{G}{$\Gamma$}
\psfrag{t}{$\tau$}
\begin{center}
\fbox{\epsfig{file=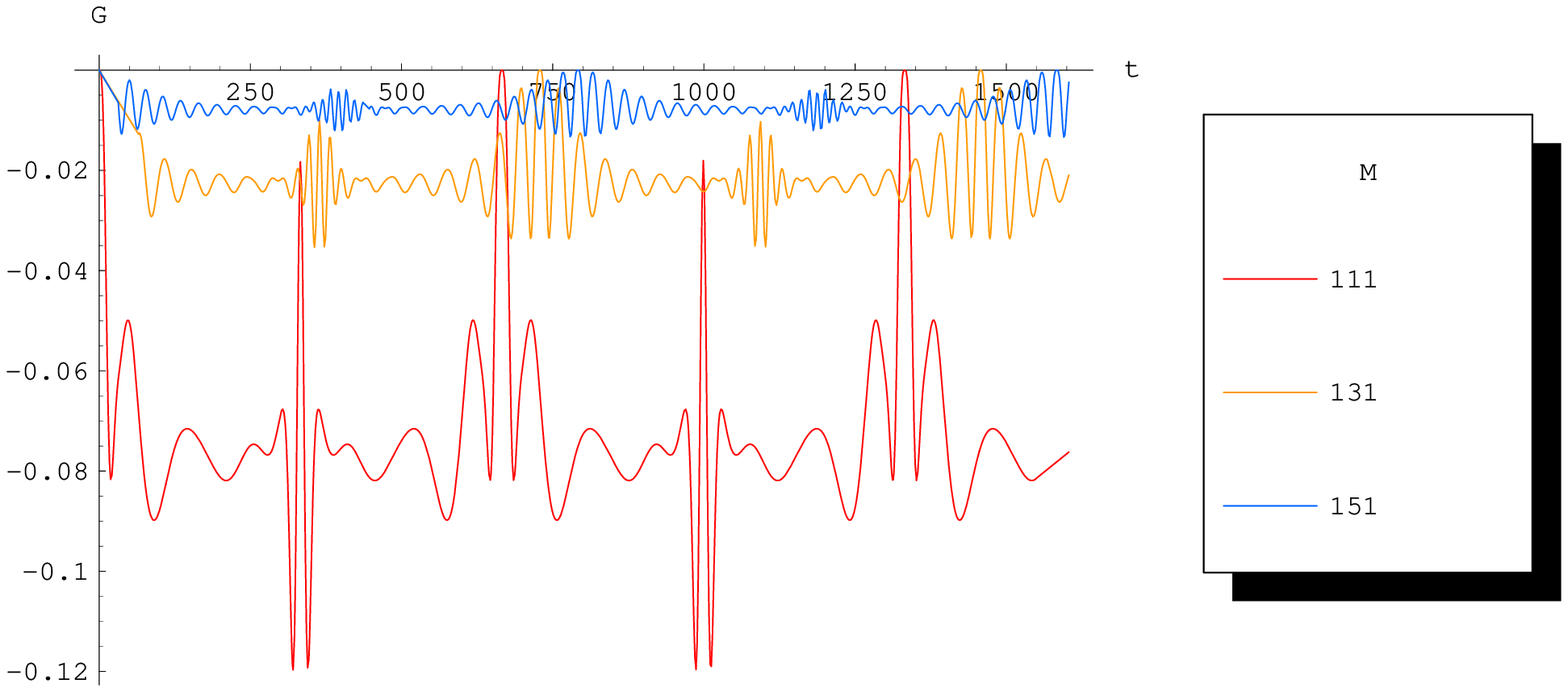,scale=0.4}}
\caption{A plot of the correlation function, $\Gamma$, using the initial conditon  in which particle 1 has isospin-up and particle 2 has isospin-down.  In these plots $N=101$ and M is varied such that $N < M$ .  The time is scaled as usual, $\tau =gt(N+M)$.  Note that the correlation function is negative for all cases.}
\label{fig:down b}
\end{center}
\end{figure}

In this case the density matrix of \eq{twoBodyDensityMatrix} is used.  Figure \ref{fig:down a} and figure \ref{fig:down b} show plots of the correlation function for $N=101$ and various numbers of M.

The features of this correlation function are similar to those found in the correlation function in the case of initially parallel isospins, and also by Friedland, McKellar and Okuniewicz  for single particle probabilities.  That is, the correlation function displays equilibration on an incoherent timescale, and also freeze-out and periodicity.  In the case $M \sim N$, the correlation is again of order 10\%.  A difference from the parallel isospin case is that the correlation function is negative. 

Our study of the two body correlations in this solvable model show that, in precisely the case when the equilibration occurs on an incoherent timescale, the correlations are not small.  This leads us to ask the question:
\begin{quotation} \emph{
Are the correlations big enough to induce significant corrections to the single particle kinetic equations, in spite of the incoherent time evolution?}
\end{quotation}
Since we are working with a solvable model, and thus we know the time evolution of the complete many particle density matrix, we can proceed to answer this question.

\section{\label{sec:BBGKY}Correlations and the time development of the single particle density matrix}
\subsection{The BBGKY Equation In the  Model}
\label{sec:ChKineticEq}

The  time evolution of the density matrix, is described by the Von Neumann equation,
\begin{equation}
	i \frac{\partial \rho}{\partial t}=[H,\rho].
\end{equation}
We can split the Hamiltonian into its single particle and interaction constituents,
\begin{equation}
	H=\sum^N_{i=1}{K_i}+\frac{1}{2}\sum_{i\neq j}V_{ij},
\end{equation}
where $K$ is the single particle Hamiltonian of the $i$th particle and $V_{ij}$ is the interaction Hamiltonian between the $i$th and $j$th particle.

The rate of change of a single particle density matrix is found by tracing over all the other particles,
\begin{equation}
	\begin{split}
		i \frac{\partial \rho^{(1)}_1}{\partial t}&=[K_1,\rho^{(1)}_1]+ \frac{1}{2}\sum_{i\neq j}\mbox{Tr}_{2..N}[V_{ij},\rho_{1..N}]\\
		&=[K_1,\rho^{(1)}_1]\\
		& \quad+\frac{1}{2}\sum_j \mbox{Tr}_j[V_{1j},\rho^{(2)}_{1j}]+\frac{1}{2}\sum_i \mbox{Tr}_i[V_{i1},\rho^{(2)}_{1i}]\\
		 &\quad+\frac{1}{2}\sum_{i\neq j} \mbox{Tr}_{ij}[V_{ij},\rho^{(3)}_{1ij}]\\
		&=[K_1,\rho^{(1)}_1]+\sum_j \mbox{Tr}_j[V_{1j},\rho_{1j}]\\
		&\quad+\frac{1}{2}\sum_{i\neq j} \mbox{Tr}_{ij}[V_{ij},\rho^{(3)}_{1ij}].
	\end{split}
\label{eq:bbgky}	
\end{equation}
In the last line of \eq{eq:bbgky} we have employed the symmetry of our interaction Hamiltonian.

\eq{eq:bbgky} is the first quantum  Bogolubov-Born-Green-Kirkwood-Yvon (BBGKY) equation\footnote{The quantum BBGKY equations have a long history, see for example\cite{Snider1960, Orlov1988}, but the derivation simply follows \emph{mutatis mutandis} from that of the classical BBGKY equations.  Chapman and Cowling\cite{cc70} give the classical equations with a summary of their history (Yvon actually derived them first), and Huang\cite{Huang87} gives a useful discussion of them.}.

In the Friedman-McKellar-Okuniewicz model there is no single particle Hamiltonian.  In this general discussion we retain such a Hamiltonian, which is responsible for the free space neutrino oscillations, but in numerical examples it is omitted, although it can be introduced with just a little extra effort.  We will frame this general discussion for the $\mathcal{N}$ flavor case.

For the interaction Hamiltonian of our model it can readily be shown that the last term goes to zero, and the three particle density matrix does not contribute to the time dependence of the single particle density matrix.  This is a significant simplification in the dynamics of the model.
To demonstrate this represent the one, two and three particle operators on the flavor space on the basis of direct products of the extended Gell-Mann matrices $\lambda_\mu$, $\mu \in \{0, 1, \ldots, \mathcal{N}^2 - 1\}$, which were introduced above.  The superscript ${}^{[i]}$ indicates that the Pauli martix or function refers to particle $i$, and similarly for multiple bracketed superscripts.  Explicitly
\begin{eqnarray}
K_i & = & \half \mathcal{K}^{[i]}_\mu \lambda^{[i]}_\mu, \\
V_{ij}  & = & \quarter v^{[ij]}_{\mu\nu} \lambda^{[i]}_\mu \otimes \lambda^{[j]}_\nu, \\
\rho^{(1)}_i & = &\half  \mathcal{P}^{[i]}_\mu \lambda^{[i]}_\mu. \\
\rho^{(2)}_{ij} & = & \quarter p^{[ij]}_{\mu\nu}\lambda^{[i]}_\mu \otimes \lambda^{[j]}_\nu. \\
\rho^{(3)}_{ijk} & = &\eighth \pi^{[ijk]}_{\kappa\mu\nu} \lambda^{[i]}_\kappa \otimes \lambda^{[j]}_\mu \otimes \lambda^{[k]}_\nu.
\end{eqnarray}
Here the superscript $^{[i]}$ refers to particle $i$, and we have employed a summation convention over repeated Greek indices.  
The few particle density matrices satisfy 
\begin{eqnarray}
\Tr(\rho^{(1})_i & = &  1\\
\Tr_j(\rho^{(2)}_{ij}) & = & \rho^{(1)}_i .\label{eq:rho12}
\end{eqnarray}
Note that \eq{eq:rho12} is valid for all $N-1$ possible values of $j$.  From these follow  the relationships
\begin{eqnarray}
 \mathcal{P}^{[i]}_0 & = & 1, \\
 p^{[ij]}_{00} & = & 1,\\
 p^{[ij]}_{\mu 0} & = &   \mathcal{P}^{[i]}_\mu, \\
 p^{[ij]}_{ 0 \nu} & = &  \mathcal{P}^{[j]}_\nu.
 \end{eqnarray}
 
 The algebra of the extended Gell-Mann matrices is
\begin{eqnarray}
\left\{\lambda_\mu, \lambda_\nu\right\} & = &  2 d_{\mu\nu\kappa}\lambda_\kappa \\
\left[\lambda_\mu, \lambda_\nu\right] & = & 2 if_{\mu\nu\kappa}\lambda_\kappa \\
\lambda_\mu \lambda_\nu & = &  d_{\mu\nu\kappa}\lambda_\kappa+  if_{\mu\nu\kappa}\lambda_\kappa.
\end{eqnarray}
The coefficients $d_{\mu\nu\kappa}$ defined  from the anticommutators are completely symmetric in their indicies, and the coefficients $f_{\mu\nu\kappa}$ defined  from the commutators are completely antisymmetric.  These coefficients may, if required explicitly, be calculated from
\begin{equation}
d_{\mu\nu\kappa} = \frac{1}{4} \mbox{Tr}\left(\left\{\lambda_\mu, \lambda_\nu\right\} \lambda_\kappa\right) \quad\quad 
f_{\mu\nu\kappa} = \frac{-i}{4} \mbox{Tr}\left(\left[\lambda_\mu, \lambda_\nu\right] \lambda_\kappa\right).
\end{equation}
Using this algebra, the components of the final term in \eq{eq:bbgky} 
 may be written as
\begin{equation}
\begin{split}
 \mbox{Tr}_{ij}[V_{ij},\rho^{(3)}_{1ij}]
& = \frac{1}{32} v^{[ij]}_{\mu\nu}\pi^{[1ij]}_{\kappa\mu'\nu'}\lambda^{[1]}_{\kappa} \times \\
&
\mbox{Tr}_{ij}\left(if_{\mu\mu'\alpha}\lambda^{[i]}_\alpha \lambda^{[j]}_\nu\lambda^{[j]}_{\nu'} + \lambda^{[i]}_\mu\lambda^{[i]}_{\mu'} (i f_{\nu{\nu'}\beta} \lambda^{[j]}_\beta\right)\\
&
 = 0,
\end{split}
\end{equation}
showing that the
BBGKY equation for $\dot{\rho}^{(1)}$ depends only on $\rho^{(1)}$ and $\rho^{(2)}$.
\begin{equation}
	\begin{split}
		i \frac{\partial \rho^{(1)}_1}{\partial t}&=\\
		&[K_1,\rho^{(1)}_1]+\sum_j \mbox{Tr}_j[V_{1j},\rho^{(2)}_{1j}].	\end{split}
\label{eq:bbgky2}	
\end{equation}
In terms of the Gell-Mann matrix representation of the density matrices, \eq{eq:bbgky2} is 
\begin{equation}
\dot{\mathcal{P}}^{[i]}_\mu = f_{\alpha\beta\mu}\left(\mathcal{K}^{[i]}_\alpha\mathcal{P}^{[i]}_\beta + \sum_j v^{[ij]}_{\alpha\lambda}p^{[ij]}_{\beta\lambda}\right).
\end{equation}
Note that, in the general case, $\mathcal{P}^{[i]}_0$ is constant, as it must be, but that $\vec{\mathcal{P}^{[i]}}^2$ is not constant. Thus in general the one particle system is not a pure state, in that $(\rho^{[1]})^2 \ne \rho^{[1]}$.
We have an explicit representation of the one and two body density matrices, in the case that $K_j = 0$, from the previous section, and it is a good check on our calculations to verify that the \eq{eq:bbgky2}	is indeed satisfied.  This has been done, and as the calculation is rather involved, the interested reader is referred to the Honours thesis of JQ \cite{Quach:2009vh} for the details.

\subsection{\label{sec:Boltz}The Boltzmann equation and its failure in the model}

While it is reassuring that the exact BBGKY equation is satisfied in our exactly solvable model, we really need to test the analogue in the model of the neutrino kinetic equations.  As shown in detail by Thomson\cite{Thomson:1990th} and McKellar and Thomson\cite{McKellar:1992ja} there are two key approximations in deriving the neutrino kinetic equations.
\begin{itemize}
\item Two particle correlations can be neglected.
\item The time of a collision is negligible compared to the time between collisions.
\end{itemize}
In the present model, as we have no spatial dependence in the interactions we cannot model the second approximation.  The first approximation can be implemented in the model by approximating the two body density matrix as the direct product of single particle density matrices, 
\begin{equation}
\rho^{(2)}_{ij} \approx \rho^{(1)}_i \times \rho^{(1)}_j,
\end{equation}
and substituting this approximation in \eq{eq:bbgky2}.  In this way we obtain a version of the quantum Boltzmann equation
\begin{equation}
\dot{\mathcal{P}}^{[i]}_\mu = f_{\alpha\beta\mu}\left(\mathcal{K}^{[i]}_\alpha +  \sum_{j \ne i} v^{[ij]}_{\alpha\lambda}{\mathcal{P}^{[j]}_\lambda}\right){\mathcal{P}^{[i]}_\beta}. \label{b-1}
\end{equation}
Before we specify the interaction appropriate to the model, we see that in the Boltzmann equation \eq{b-1} the interactions now give a modified oscillatory behaviour, as $\vec{\mathcal{P}}^{[i]}$ ``precesses'' about the vector with components $\mathcal{K}^{[i]}_\alpha + \sum_{j \ne i} v^{[ij]}_{\alpha\lambda}{\mathcal{P}^{[j]}_{\lambda}}$, and thus $\left(\vec{\mathcal{P}}^{[i]}\right)^2$ is a constant.  which  can be fixed to 1.  With the general result that $\mathcal{P}_0^{[i]} = 1$ this implies that $(\rho^{(1)})^2 = \rho^{(1)}$.  In the Boltzmann approximation the single particle density matrix represents a pure state.

Now we note that in the model, the Gell-Mann matrix representation of the interaction is
\begin{eqnarray} 
v_{\alpha\beta}  & = &  \phi_\alpha \delta_{\alpha\beta} \quad \mbox{with} \quad \phi_0 = 4g(\mathcal{N}+1)/\mathcal{N}  \nonumber \\
&&  \mbox{and} \quad \phi_\alpha = 2g \quad \mbox{for} \quad \alpha \ne 0,
\end{eqnarray}
 so the model Boltzmann equation is 
\begin{equation}
\dot{\mathcal{P}}^{[i]}_\kappa =  f_{\alpha\beta\kappa} \left(\mathcal{K}^{[i]} + 2 g\sum_{j} \mathcal{P}^{[j]}\right)_\alpha \mathcal{P}^{[i]}_\beta, \label{b-2}
\end{equation}
where the restriction $j \ne i$ is no longer necessary as the contribution from $j = i$ vanishes.
From this it follows that 
 $\sum_j\vec{\mathcal{P}}^{[i]} $ ``precesses'' freely, {\textit i.e.} undergoes free oscillations, and for the case studied above in which $\mathcal{K}^{[i]}$ vanishes, this vector is a constant.
 
Further, if we specify the initial state so that all particles at $t=0$ are in flavor eigenstates, as was done for $\mathcal{N} = 2$ above, 
\begin{eqnarray}
\mathcal{P}^{[i]}_\alpha(t = 0) & = &  0 \quad \mbox{unless} \quad \alpha = n^2 - 1, \nonumber \\
&&  \mbox{for} \quad n \in \{1, \ldots, \mathcal{N}\}.
\end{eqnarray}
Thus at $t = 0$ 
\begin{equation}
\left. \frac{d}{dt} \mathcal{P}^{[i]}_\beta \right|_{t = 0} = 0,
\end{equation}
and so at the slightly later time $\delta t$,
\begin{equation}
 \mathcal{P}^{[i]}_\alpha(t = \delta t) = 0 \quad \mbox{unless} \quad \alpha = n^2 - 1.
 \end{equation}
 Thus
\begin{equation}
 \frac{d}{dt} \mathcal{P}^{[i]}_\beta = 0 \quad \mbox{at all times}.
\end{equation}

This is precisely the result obtained by explicit calculation\cite{Quach:2009vh} in the case $\mathcal{N} = 2$, which 
 is in complete contradiction with the results of Friedland, McKellar and Okuniewicz\cite{Friedland:2006ke}, in which the exact non-trivial time development of $\rho^{(1)}$ was computed.  The Boltzmann equation is not satisfied in the model in the most dramatic way possible --- \textbf{all of the time dependence of $\rho^{(1)}$ is generated by the correlations} --- even though the correlation function, as we saw, has a magnitude of at most about 10\%.

 Clearly the Boltzmann equation is not even approximately valid in this model.

\section{Conclusion}

It is tempting to suggest that the breakdown of the Boltzmann equation in this exactly solvable model is closely connected to the structure of the model, and is not a general effect.  In part the result is a consequence of the initial conditions.  Even for $\mathcal{N} = 2$, if we prepare some of the particles initially in an eigenstate of, say, $t_x$ or the mass operator, and others (the flavor eigenstates) in  eigenstates of $t_z$,  then the rhs of the Boltzmann equation is non-vanishing.  Without further calculations it is not possible to say how well the Boltzmann equation is satisfied, but at least it is not completely and dramatically wrong.  The way in which one can obtain a non-vanishing time dependence of the one particle density matrix from the Boltzmann equation is to have some of the initial neutrinos as one class of eigenstates, and the rest as another class of eigenstates.  It is not easy to imagine how nature may have prepared such a state.

The result is specific to the model interaction, and that interaction is not particularly realistic.  An outstanding challenge is  to generalize the interaction, maintaing the solvability of the model, and testing the Boltzmann equation in this more appropriate model.

Regrettably, we have found that, while a coherent time scale may indicate the importance of correlations, an incoherent timescale is not a reliable indicator of the absence of significant correlations.  As a consequence this particular solvable model is not a good guide to the validity or otherwise of the neutrino kinetic equations in either the early universe or supernovae.

\section*{Acknowledgements}

This work has been supported in part by the Australian Research Council.

\bibliography{kinetic}

\end{document}